# Structural and optical studies of FeSb$_2$ under high pressure

Claudio Michel Poffo, Sergio Michielon de Souza[*], and Daniela Menegon Trichês[*]

*Departamento de Engenharia Mecânica, Universidade Federal de Santa Catarina, Campus Universitário Trindade, S/N, C.P. 476, 88040-900 Florianópolis, Santa Catarina, Brazil*

João Cardoso de Lima and Tarciso Antonio Grandi

*Departamento de Física, Universidade Federal de Santa Catarina, Campus Universitário Trindade, S/N, C.P. 476, 88040-900 Florianópolis, Santa Catarina, Brazil*

Alain Polian and Michel Gauthier

*Physique des Milieux Denses, IMPMC, CNRS-UMR 7590, Université Pierre et Marie Curie-Paris 6, B115, 4 Place Jussieu, 75252 Paris Cedex 05, France*

**ABSTRACT**

Nanostructured orthorhombic FeSb$_2$ and an amorphous phase were formed by mechanical alloying starting from a mixture of high purity elemental Fe and Sb powders. The effects of high pressures on structural and optical properties were studied using X-ray diffraction (XRD) and Raman spectroscopy (RS). XRD patterns showed the presence of the orthorhombic FeSb$_2$ phase up to the maximum pressure applied (28.2 GPa). The XRD patterns showed also an increase in the amount of the amorphous phase with increasing pressure up to 23.3 GPa. At 14.3 GPa, together with the former phases, a new phase was observed and indexed to a tetragonal FeSb$_2$ phase, but its volume fraction is small at least up to 23.3 GPa. For the orthorhombic FeSb$_2$ phase, the pressure dependence of the volume fitted to a Birch-Murnaghan equation of state gave a bulk modulus $B_0$ = 74.2 ± 3.0 GPa and its pressure derivative $B'_0$ = 7.5 ± 0.6. RS measurements were performed from atmospheric pressure up to 45.2 GPa. For the orthorhombic FeSb$_2$ phase, the Raman active $A_g^2$ mode was observed up to the maximum pressure applied, while the $B_{1g}^1$ mode disappeared at 16.6 GPa. For pressures higher than 21 GPa, the Raman active $E_g^2$ mode of a tetragonal FeSb$_2$ phase was observed, confirming *ab initio* calculations reported in the literature.

## I. INTRODUCTION

FeSb$_2$ is a narrow-gap semiconductor that has attracted a lot of attention because of its unusual magnetic properties (paramagnetic to diamagnetic crossover at around 100 K), thermoelectric properties (colossal Seebeck coefficient $S$ at 10 K and the largest power factor $S^2\sigma$ ever reported),[1] and transport properties (a metal-to-semiconductor transition at around 40 K).[2]

The FeSb$_2$ compound can be synthesized by the self-flux method[3] and the high-temperature flux method,[4] among others. These techniques, besides being expensive, do not allow a good control of the size of the nanoparticles. At room temperature and atmospheric pressure, FeSb$_2$ crystallizes in an orthorhombic structure (s.g. Pnnm, Z = 2), with Fe atoms at the 2$a$ (0, 0, 0) Wyckoff position and Sb atoms at the 4$g$ (x, y, 0) position, where x = 0.1885 and y = 0.3561. Each Fe atom has a deformed octahedral environment, and octahedra share edges along the c axis.[5]

Nanostructured materials have been widely studied due to their interesting properties. For example, the performance of a thermoelectric material can be improved if its thermal conductivity is reduced without strong degradation of its electrical properties. The dimensions of crystallites of nanostructured materials may allow an important reduction in the thermal conductivity of the lattice and promoting an improvement in their thermoelectric conversion efficiency. Nanostructured materials have been produced by different techniques, including mechanical alloying (MA). Suryanayarana´s paper[6] gives a good review of the MA technique, while the involved physical mechanisms are described in Refs. 7–10.

From the structural point of view, nanostructured materials have two components: crystallites of nanometric dimensions 2–100 nm, having the same structure as their crystalline counterparts, and an interfacial component, formed by different types of defects (grain boundaries, interphase boundaries, dislocations, etc.) that surround the crystallites. Nanostructured materials are metastable.[11]

The literature describes studies on the effect of high pressure in several nanostructured materials,[12-18] but the nanostructured FeSb$_2$ alloy is not among them. Petrovic et al.[19] investigated the effect of high pressure up to 7.14 GPa on bulk orthorhombic $FeSb_2$ compound and no structural changes were observed. The compounds $TiSb_2$, $VSb_2$, $MnSn_2$, $CoSn_2$ and $FeSn_2$ crystallize in the $CuAl_2$ structure (s.g. I4/mcm, Z=4).[5] Takizawa *et al.*[20] reported a $CrSb_2$ phase that crystallized in the $CuAl_2$ structure above 5.5 GPa. Wu *et al.*[21] using *ab initio* calculations predicted a transformation of orthorhombic $FeSb_2$ phase into a tetragonal structure at 38 GPa.

For some years, we focused our research in studying the effect of high pressure in nanostructured materials, mainly those with thermoelectric applications.[12-14,18,22] Due to the scientific interest in $FeSb_2$ and its technological importance, we investigated the effect of high pressure on nanostructured $FeSb_2$ powder produced by the MA technique. This study, besides filling a gap in the literature, will extend the investigation started by Petrovic *et al.* [19] on the effect of high pressures on this compound. This paper reports the results of this study.

## II. BRIEF CONSIDERATIONS ABOUT THE FORMATION OF CRYSTALLINE OR AMORPHOUS BINARY ALLOYS AT HIGH PRESSURES AND TEMPERATURES

In order to understand the formation of crystalline and amorphous phases at high pressures and temperatures, we will consider the simplest case, i.e., that of binary alloys. Besides other physical mechanisms, there are two key points to taken into account: (i) the two elements must have a large negative relative heat of mixing, and (ii) either one of them is an anomalously fast diffuser (leading to an amorphous phase), or the two elements have similar diffusion coefficients (leading to a crystalline phase).

When high pressure is applied in a binary mixture of high purity elemental powders sealed in a diamond anvil cell (DAC), defective chemical bonds are formed, characterized by angle and length changes. This chemical disorder stores a sizable amount of energy. As the pressure is increased further, a composite powder containing particles with defective chemical bonds is formed. When the material is heated to an appropriate temperature, the total energy (thermal energy plus the energy released from the defective chemical bonds) becomes the driving force for atom diffusion, promoting solid state reactions and resulting in the formation of crystalline and/or amorphous phases. Usually, the crystalline phase is the same that is produced through conventional methods and is thermodynamically stable at room temperature and atmospheric pressure, but new phases can be obtained through a convenient choice of pressure and temperature. For example, Takizawa *et al.*[20] produced the orthorhombic $CrSb_2$ phase

(stable at room temperature and atmospheric pressure) starting from a $CrSb_2$ mixture and using pressures smaller than 5.5 GPa and temperatures below 1073 K. For pressures greater than 5.5 GPa and temperatures above 773 K, a new tetragonal $CrSb_2$ phase was obtained.

When high pressure is applied to a single crystal or bulk compound sealed in a DAC, chemical disorder is induced and energy is stored, as was previously described. If the pressure is increased further, two things may happen: 1) the defective chemical bonds will break, releasing the stored energy, and 2) the defective chemical bonds will not break. When the defective chemical bonds break and the material is heated to an appropriate temperature, the total energy (thermal energy plus the energy released from the defective chemical bonds) becomes the driving force to promote the diffusion of free atoms, resulting in the formation of one or more new crystalline and/or amorphous phases. When the defective chemical bonds do not break, only a gradual topological rearrangement of chemical bonds occurs to minimize the energy and a new crystalline phase is formed. In this case, appropriate heating can accelerate the formation of the new phase and decrease the pressure at which phase transformation occurs. The driving force to promote the topological rearrangement of chemical bonds is the sum of thermal energy and the energy released from defective chemical bonds. If the material is not heated, the driving force to promote the topological rearrangement is lower and the formation of a new phase occurs at a higher pressure. The work reported by Takizawa *et al.*[20] is a good example of no breaking of defective chemical bonds. These researchers used temperature and pressure to accelerate the topological rearrangement of chemical bonds and decrease the pressure of transformation from bulk orthorhombic $CrSb_2$ phase into a tetragonal $CrSb_2$ phase. This new phase was formed using pressures above 5.5 GPa and temperatures above 773 K. Other work reported by Nakayama *et al.*[23] is an example of breaking the defective chemical bonds and forming a mixture of phases. These researchers applied high pressure to bulk rhombohedral $Bi_2Te_3$ at room temperature and reported the formation of a mixture of phases for pressures between 9.2 and 16.2 GPa.

When high pressure is applied to binary nanostructured compounds sealed in a DAC, a behavior different of those previously described is observed. This happens because these materials are structurally formed by two components: crystallites with dimensions < 100 nm, with a structure similar to their bulk counterparts, and an interfacial region, formed by different kinds of defects, that surround the crystallites.

Both components have similar number of atoms, but the atoms are distributed in different atomic arrangements in the interfacial component. Thus, the stored energy is larger in the second component. The pressure increase affects first the interfacial component, promoting a gradual elimination of defects and a release of the stored energy. In addition, the atoms of this component located near the interfaces are incorporated in the crystallites, promoting their growth. Only after this process is over the pressure affects the crystalline component, removing strains. Two consequences are easily seen: (1) an improvement in crystallinity, and (2) an increase in value of pressure at which occurs transformation of atmospheric pressure phase to high pressure phase, when compared that of its bulk counterpart. In the case of a transformation from a high pressure phase to another, this effect is not observed, i.e., phase transformations do not occur at lower pressures in the nanosctructures material. For example, Trichês et al.[22] investigated the effect of high pressure on the nanostructured orthorhombic ZnSb phase and observed a phase transformation into an hP1 phase between 11 and 14.6 GPa, while the literature reports a transformation of bulk orthorhombic ZnSb phase into the same phase at 7 GPa.[24] No heat treatment was performed.

Due to the fact that the interfacial component consists of different kinds of defects, different regions with different compositions can be present. If there are no strong differences among the enthalpies of formation of phases corresponding to different compositions, more than one phase can be nucleated when pressure is applied. An example will be shown in this study, where the microstructure of the as-milled sample consists of nanostructured orthorhombic $FeSb_2$ and an amorphous phase. With increasing pressure, nucleation of a new phase as well as an increase in the volume fraction of the amorphous phase were observed, without structural degradation of the orthorhombic phase.

**III. EXPERIMENTAL PROCEDURE**

A binary $FeSb_2$ mixture of elemental powders of Fe (Aldrich, purity 99.999%) and Sb (Alfa Aesar, purity 99.999%) was sealed together with several steel balls of 11.0 mm in diameter into a cylindrical steel vial under argon atmosphere. The ball-to-powder weight ratio was 7:1. The vial was mounted on a SPEX Mixer/mill, model 8000. The temperature was kept close to room temperature by a ventilation system. The structural changes of the mixture as a function of milling time were followed by XRD measurements. After 11 hours of milling, the XRD pattern showed an excellent agreement with that given in the ICSD Database[5] (code 41727) for the orthorhombic FeSb$_2$ phase (s.g. Pnnm, Z=2). The milling process was extended to 32 hours but no further structural changes were observed.

A DAC with an opening that allowed probing up to 2θ = 28° was used.[25] A small amount of FeSb$_2$ powder was compacted between diamonds to a final thickness of approximately 15 µm. A small chip of this preparation, about 80 µm in diameter, was loaded into a stainless-steel gasket hole of 150 µm diameter. Neon gas was used as a pressure-transmitting medium because (i) it is one of the softest materials, (ii) it is chemically inert, and (iii) it has no luminescence and no Raman activity. The pressure was determined by the fluorescence shift of a ruby sphere loaded in the sample chamber.[26] The quasi-hydrostatic conditions were controlled throughout the experiments by monitoring the separation and widths of $R_1$ and $R_2$ lines. *In situ* XRD patterns as a function of pressure were acquired at the XRD1 station of the ELETTRA synchrotron radiation facility. This diffraction beamline is designed to provide a monochromatized, high-flux, tunable x-ray source in the spectral range from 4 to 25 keV.[27] The study was performed using a wavelength of 0.068881 nm (E = 18,002.06 eV). The detector was a 345-mm imaging plate from MarResearch. The sample-to-detector distance was calibrated by diffraction data from Si powder loaded in the diamond anvil cell. The XRD data were collected at 0.7, 2.2, 3.7, 5.3, 6.6, 8.1, 9.7, 12.6, 14.3, 15.6, 17.6, 19.1, 21.0, 23.3 and 28.2 GPa. An exposure time of 10 min was used for all measurements. The two-dimensional diffraction patterns were converted to intensity versus 2θ using the fit2D software[28] and analyzed by the Rietveld method[29] using the GSAS package.[30]

For the Raman measurements as a function of pressure, a particle of approximately 50 x 60 x 20 µm$^2$ was loaded in the DAC. The Raman spectra and ruby luminescence were recorded in the backscattering geometry by means of a Jobin-Yvon T64000 Raman triple spectrometer and a liquid-nitrogen-cooled charge coupled device multichannel detector. An excitation line of λ = 514.5 nm of an Ar laser was used for excitation and focused down to 5 µm with a power of about 20 mW at the entrance of the DAC. The Raman spectra were collected at 1.3, 5.3, 7.5, 9.6, 12.7, 16.6, 21.0, 24.9, 28.3, 32.4, 36.9, 42.7 and 45.2 GPa. An exposure time of 120 min was used for all measurements. The Raman frequencies were determined from a fit of the peaks to a Lorentzian profile. The frequency accuracy was better than 1 cm$^{-1}$.

## IV. RESULTS AND DISCUSSION

### A. XRD pattern at room temperature and atmospheric pressure

For a milling time of 32 hours, the XRD pattern showed an excellent agreement with that given in the ICSD Database[5] (code 41727) for the orthorhombic $FeSb_2$ phase (s.g. Pnnm, Z=2). In addition, diffraction peaks of unreacted elemental Sb and a broad background under the most intense peaks of the orthorhombic $FeSb_2$ phase were observed. In order to investigate the origin of this background, a small amount of the as-milled powder was analyzed by the differential scanning calorimetry technique. The first measurement showed an intense exothermic peak located at about 611 K, while in the second this peak was absent. Thus, it was concluded that besides the orthorhombic $FeSb_2$ phase the milling process generated an amorphous phase. Petrovic *et al.*[19] using the levitation melting technique produced single crystal of orthorhombic $FeSb_2$ together with a small content (8%) of unreacted elemental Sb. These results suggest that to obtain a pure $FeSb_2$ phase a compensation of elemental Sb should be used. In this study, no compensation was made.

For 32 hours of milling, the XRD patterns at atmospheric pressure and at 0.7 GPa were similar. In order to decrease the number of figures to be presented in this paper, only the last will be shown. Another paper reporting these results, among others, is in course.

All the peaks observed on the XRD pattern of the orthorhombic $FeSb_2$ phase have their base enlarged, suggesting that the as-milled sample has crystallites with very small dimensions. The mean size of the crystallites was estimated using the equation below, which takes into account the line broadening caused by both crystallite size and strain.[31]

$$\left(\frac{\beta_t \cos\theta}{K\lambda}\right)^2 = \frac{1}{d^2} + \sigma_p^2 \left(\frac{\sin\theta}{K\lambda}\right)^2 \qquad (1)$$

Here, $\theta$ is the diffraction angle, $\lambda$ is the X-ray wavelength, $\beta_t$ is the total broadening measured at the full-width at half-maximum (FWHM) of the peak in radians, $d$ is the crystallite size, $\sigma_p$ is the strain, and $K$ is a constant dependent on the measurement conditions and on the definition of $\beta_t$ and $d$ (here $K$ was assumed to be 0.91 as is usual in the Scherrer formula). Graphical linearization of the above relationship, *i.e.*, plotting $\beta_t^2 \cos^2\theta/\lambda^2$ versus $\sin^2\theta/\lambda^2$, yields the mean crystallite size free from strain effects from the values of the intercept of the straight line obtained, as well as the strain obtained from the slope. From the Rietveld refinement, the generated $\beta_t$ and $2\theta$ positions values were used in Eq. (1), and the calculated mean crystallite size and strain were d ≈ 26 nm and σp ≈ 0.4 %. The mean crystallite size shows that the as-milled $FeSb_2$ powder has a nanometric structure. From the Rietveld refinement, 7 % of unreacted elemental Sb was calculated.

Recently, we developed an approach to estimate the volume fractions of crystalline and interfacial components.[32,33] and it was used to estimate the contribution of nanometric $FeSb_2$ crystallites to the XRD pattern of the as-milled powder. For this, the measured intensity was corrected for polarization and reabsorption, and converted to electron units using the mean square scattering factor <$f^2$> of $FeSb_2$ analytically calculated. After that the inelastic scattering was subtracted. The contribution of the interfacial component to the XRD pattern is diffuse. When an amorphous phase is present, its contribution overlaps that of the interfacial component, making impossible to distinguish the contributions of each. An evaluation of the interfacial component contribution to the normalized XRD pattern and its subtraction yield the contribution of nanometric $FeSb_2$ crystallites. The interfacial component contribution was estimated using the Origin software.[34] The ratio of the integrated intensity of the contribution of nanometric $FeSb_2$ crystallites to that of the whole XRD pattern yields a crystalline volume fraction of ≈ 63% and, consequently, the interfacial plus amorphous volume fraction is ≈ 37%.

**B. High pressure XRD measurements**

As mentioned previously, the XRD patterns measured at atmospheric pressure and at 0.7 GPa are similar. The former was indexed to the orthorhombic $FeSb_2$ phase. Figures 1 and 2 show *in situ* synchrotron XRD patterns of the nanostructured $FeSb_2$ sample for several pressures up to 28.2 GPa. The initial XRD pattern is seen up to the highest pressure used. With increasing pressure, the diffraction peaks shift toward higher $2\theta$ values and those initially located between $2\theta$ = 13.5° and 16° become well separated. The intensity of these peaks decreases up to 12.6 GPa and increases for higher pressures. No XRD measurements were performed as the pressure was decreased. As mentioned in the previous section, 7 % of unreacted elemental Sb is present in as-milled powder. In Fig.1, at 0.7 GPa, these diffraction peaks are marked with an asterisk (*) symbol and are observed up to 9.7 GPa. In all XRD patterns, between $2\theta = 13.5°$ and $17°$, one can see an amorphous phase which reaches maximum intensity at 23.3 GPa. There is no structural degradation of the orthorhombic $FeSb_2$ phase with increasing pressure. This indicates that the interfacial component is responsible for the increase in the volume fraction of the amorphous phase content, and an explanation is the following: at atmospheric pressure, the volume fraction of the interfacial component and amorphous phase is $\approx 37$ %. Increasing the pressure promotes compaction, elimination of defects and release of stored energy. The free Fe and Sb atoms diffuse and are incorporated by the amorphous phase.

We used the software DATLAB code[35] to estimate the volume fraction of the amorphous phase at 6.6, 14.3, 23.3 and 28.2 GPa, as shown in Fig. 3. The intensity of main halo is maximum at 23.3 GPa. The increase in diffuse scattering for angles larger than $2\theta = 18°$ can be attributed to the Fe $K_\alpha$ and/or $K_\beta$ fluorescence generated during the measurements, since the photon energy was 18,002.06 eV and the Fe $K$-edge is 7112 eV. We used the Ehrenfest relation $r = \lambda/{E\sin\theta}$,[36] where the structure dependent constant E was taken to be 1.671 and $\lambda$ is the wavelength used in the experiments, to estimate the interatomic distance for the first neighbors. The values were 3.20, 3.12, 3.05 and 2.95 Å for 6.6, 14.3, 23.3 and 28.2 GPa, respectively. These values decrease with increasing pressure, as expected, and are compatible with the interatomic distances found in the Fe-Sb alloys.

The peak at about $2\theta = 20.8°$ in the XRD pattern for 9.7 GPa, which was also seen in the patterns for 14.3, 15.6 and 19.1 GPa, is attributed do the gasket, confirmed by patterns taken without any sample.

At 14.3 GPa, besides the diffraction peaks of the orthorhombic phase, other low intensity diffraction peaks were observed at about $2\theta = 11.5°$ and $18.4°$, indicating the presence of a new phase. As there is no structural degradation of the orthorhombic and amorphous phases with increasing pressure, it is concluded that this new phase was nucleated in the interfacial component. In the same XRD pattern there are two diffraction peaks at about $2\theta = 19.7°$ and $22.8°$ that were indexed to (111) and (200) planes of neon (N), which crystallizes in a f.c.c structure at about 10 GPa.[37] In this study, neon gas was used as pressure-transmitting medium.

The synchrotron XRD patterns displayed in Figs. 1 and 2 for the orthorhombic $FeSb_2$ phase were refined using the Rietveld method,[29,30] and the results are summarized in Table I. Fig. 4 shows the experimental and simulated XRD patterns at 0.7 GPa. From the Table I one can see that the pressure dependence of the lattice parameters shows two linear behaviors denoted as regions I (from 0.7 up to 8.1 GPa) and II (from 12.6 up to 28.2 GPa). In region I, change in the lattice parameters is more abrupt. Individual lattice parameter compressibilities $\beta_0^i = -(1/i)(\partial i/\partial P)_T$ (i=a,b,c) were calculated, and the values were $\beta_0^a = 0.00255(1)$ $GPa^{-1}$, $\beta_0^b = 0.00327(9)$ $GPa^{-1}$ and $\beta_0^c = 0.00479(8)$ $GPa^{-1}$ for the region I and $\beta_0^a = 0.00153(4)$ $GPa^{-1}$, $\beta_0^b = 0.00159(2)$ $GPa^{-1}$ and $\beta_0^c = 0.00199(9)$ $GPa^{-1}$ for the region II. Calculated values for the region I are almost twice those calculated for the region II. In both regions the c-axis is more compressive than the a- and b-axes. The calculated values for the region I agree quite well with those reported in Ref. 19, where the highest pressure was 7.14 GPa. A possible explanation for the differences in the two regions is the following: the microstructure of the as-milled sample includes crystalline and interfacial components plus an amorphous phase. With increasing pressure up to 8.1 GPa, the volume fraction of the interfacial component decreases due to partial elimination of the several kinds of defects and crystallite growth through the incorporation of atoms located near the boundaries of the crystallites. Thus, the effect of high pressures is to improve the crystallinity of the sample, leaving it energetically more stable[22] and more like bulk $FeSb_2$. On the other hand, from 12.6 up to 28.2 GPa (see Figs. 2 and 3) an important increase in the amount of the amorphous phase and a new crystalline phase are observed. As a structural degradation of the orthorhombic $FeSb_2$ phase is not observed, the different behavior in the individual lattice parameter compressibility is attributed to an increase in the volume fraction of the amorphous phase as well as to the presence of the new high pressure phase.

The volume as a function of pressure V(P) obtained from the Rietveld refinement for the orthorhombic $FeSb_2$ phase (see Table I) was fitted to a Birch–Murnaghan equation of state (BM EOS):[38]

$$P = \frac{3}{2} B_0 \left( X^{-7/3} - X^{-5/3} \right) \left\{ 1 + \frac{3}{4} \left( B_0' - 4 \right) \left( X^{-2/3} - 1 \right) \right\} \quad (2)$$

where X = ($V/V_0$). The value of $V_0$ = 122.269 Å$^3$ was obtained from the Rietveld refinement of XRD pattern measured at ambient temperature and atmospheric pressure. The best fit was reached by considering the values of bulk modulus $B_0$ = 74.2 ± 3.0 GPa and its pressure derivative $B'_0$ = 7.5 ± 0.6, as shown in Fig. 5. For bulk FeSb$_2$, Petrovic et al.[19] reported the values of $B_0$ = 84(3) GPa and $B'_0$ = 5(1), and their data are also shown in Fig. 5 (open stars), while Wu et al.[21] reported the values $B_0$ = 94 GPa and $B'_0$ = 4.9 and $B_0$ = 68 GPa and $B'_0$ = 5.9 for FeSb$_2$ crystallized in Pnnm and I4/mcm structures, respectively. According to Fecht[39] the bulk modulus $B_0$ for nanometric metals decreases with increasing volume fraction of the interfacial component. Thus, the difference between $B_0$ values obtained in this study and those reported in Refs. 19 and 21 can be due to the fact that our sample is nanostructured.

At 14.3 GPa, the XRD pattern shows the presence of a new high pressure phase besides those described previously. Takizawa et al.[20] reported that a new high-pressure CrSb$_2$ phase crystallized in a CuAl$_2$ structure-type crystal formed above 5.5 GPa, with metallic bond nature including the formation of Cr–Cr–Cr linear chain along the c-axis. Wu et al.[21] predicted a FeSb$_2$ orthorhombic-tetragonal phase transition at 38 GPa. In addition, it known that the TiSb$_2$, VSb$_2$, MnSn$_2$, CoSn$_2$, FeSn$_2$ and CuAl$_2$ compounds crystallize in CuAl$_2$ structure type.[5] The diffraction peaks of this new phase showed a good agreement with those given for the TiSb$_2$, VSb$_2$ and CrSb$_2$ compounds. The lattice parameters $a$ and $c$ were calculated using the classical relationship[40]

$$\frac{1}{d_{hkl}^2} = \frac{h^2 + k^2}{a^2} + \frac{l^2}{c^2}, \qquad (3)$$

where h, k and l are the Muller indexes. The peaks located at about $2\theta$ = 11.55°, 14.96° and 18.31° were indexed to the (200), (211) and (310) planes, respectively. The peak associated with the (211) plane is represented by a shoulder at the right of the peak located at about $2\theta$ = 14.78° associated to the (101) plane of the orthorhombic phase. The calculated values were $a$ = 6.8441 Å (6.555 Å) and $c$ = 5.2563 Å (5.631 Å). The numbers between parentheses are those given in the JCPDS Database for the VSb$_2$ phase (card No. 250055). Fig. 6 shows a comparison between the XRD patterns for VSb$_2$ (bottom curve) and our measurements at 14.3 GPa (top curve), where the peaks of the new FeSb$_2$ phase are observed. In order to improve the comparison, the pattern of the VSb$_2$ phase was multiplied by an arbitrary factor. The difference in the intensities of the peaks located at about $2\theta = 11.55°$ and $17.14°$ shows a texture in the new phase, which was taken into account during the Rietveld analysis.

The new high pressure FeSb$_2$ phase, with a CuAl$_2$-type structure, was observed from 14.3 to 23.3 GPa. Its XRD pattern was refined using the Rietveld method, and the results of the fits are summarized in Table II. Fig. 7 shows the experimental and simulated XRD patterns at 14.3 GPa. Of course, the orthorhombic and tetragonal FeSb$_2$ phases were used in the simulation. In this figure, the $2\theta$ range was reduced in order to show the details. Despite the small amount of the tetragonal phase, the peak located at about $2\theta = 14.78°$ can not be simulated with any accuracy if this phase is considered. The results showed that the lattice parameters $a$ and $c$ decrease linearly with increasing pressure. The interatomic distances of Fe-Fe and Fe-Sb also decrease with increasing pressure. Individual lattice parameter compressibilities were calculated, and the values were $\beta^a = 0.00105(6)$ GPa$^{-1}$ and $\beta^c = 0.00245(5)$ GPa$^{-1}$. As in the case of the orthorhombic phase, the c-axis is more compressive than the a- and b-axes. Armbruster et al.[41] investigated the compressibility of tetragonal TiSb$_2$ under high pressure up to 12 GPa and found a faster decrease of the $c$ parameter in comparison to $a$, and thus a decreasing $c/a$ ratio with increasing pressure. The results obtained in this study are consistent with their results.

**C. Raman measurements under pressure**

At room temperature and atmospheric pressure, orthorhombic FeSb$_2$ crystallizes in the $D_{2h}^{12}$ symmetry, and the Raman active modes at the $\Gamma$ point of the Brillouin zone are classified according to the irreducible representations of this point group,[42]

$$\Gamma = 2A_g + 2B_{1g} + 2B_{2g} + 3B_{3g} \qquad (4)$$

Figures 8 and 9 show the measured Raman spectra for the nanostructured FeSb$_2$ powder with increasing and decreasing pressure, respectively. As shown previously from XRD measurements, the as-milled sample contains 7 % of unreacted elemental Sb. Elemental Sb crystallizes in the $D_{3d}^5$ symmetry (s.g. $R\bar{3}m$, Z=6) and its two Raman active modes are: the $A_{1g}$ mode at 150 cm$^{-1}$ and a two-fold degenerated $E_g$ mode at 115 cm$^{-1}$.[43] In Fig. 8, these two Raman active modes are marked with an asterisk (*) symbol They are not observed for pressures higher than 5.3 GPa.

Lazarevic et al.[44] reported the Raman spectrum of the orthorhombic FeSb$_2$ phase at room temperature, and the wave numbers are: $B_{2g}$ = 90.4 cm$^{-1}$, $A_g$ = 150.7 and 153.6 cm$^{-1}$, $B_{3g}$ = 151.7 cm$^{-1}$ and $B_{1g}$ = 154.3 and 173.9 cm$^{-1}$. Thus, the peaks centered at about 163 and 187 cm$^{-1}$ on the Raman spectrum measured at 1.3 GPa and shown in Fig. 8 were attributed to the Raman active $A_g^2$ and $B_{1g}^1$ modes, respectively. With increasing pressure up to 16.6 GPa, the Raman active $A_g^2$ mode becomes sharper, while the $B_{1g}^1$ mode enlarges and disappears.

At 21 GPa, one new Raman active mode at about 233 cm$^{-1}$ emerges, while the Raman active $A_g^2$ mode of the orthorhombic FeSb$_2$ phase enlarges. With increasing pressure up to 45.2 GPa, both modes shift to higher wave numbers and the $A_g^2$ mode becomes broader and less intense, while the emergent mode becomes sharper and more intense. According to the XRD results between 14.3 and 23.3 GPa, a new tetragonal phase is present. For the isostructural compounds listed above with the CuAl$_2$ structure (symmetry $D_{4h}^{18}$), the Raman active modes at the Γ point of the Brillouin zone are classified according to the irreducible representations of this point group[42]

$$\Gamma = A_{1g} + B_{1g} + B_{2g} + 2E_g \qquad (5)$$

Armbruster et al.,[41,45] reported the Raman spectra at room temperature for the TiSb$_2$ and VSb$_2$ compounds, in form of powders and single crystals. For TiSb$_2$ powders, the wave numbers of Raman active modes are $A_{1g}$ = 172 cm$^{-1}$, $B_{1g}$ = 129 cm$^{-1}$, $B_{2g}$ = 180 cm$^{-1}$, $E_g^1$ = 83 cm$^{-1}$ and $E_g^2$ = 254 cm$^{-1}$, while for VSb$_2$ powders they are $A_{1g}$ = 166 cm$^{-1}$, $B_{1g}$ = 116 cm$^{-1}$, $B_{2g}$ = 176 cm$^{-1}$, $E_g^1$ = 103 cm$^{-1}$ and $E_g^2$ = 229 cm$^{-1}$. Except for the $E_g^1$ mode, the wave numbers of modes decrease with increasing atomic number of the transition metal. Thus, it is expected that for tetragonal FeSb$_2$ the wave numbers of active modes are smaller than for the VSb$_2$ powder. In Fig. 8 one can see that the Raman active $A_g^2$ mode of the orthorhombic FeSb$_2$ phase and the $A_{1g}$ and $B_{2g}$ of the tetragonal FeSb$_2$ phase are very close and it is difficult to separate them for pressures higher than 21 GPa. On the other hand, the Raman active $E_g^2$ mode of the tetragonal phase is well separated. In order to obtain its wave number at room temperature and atmospheric pressure, the maxima of peaks on the Raman spectra between 21 and 45.2 GPa were fitted to a linear polynomial and the presssure was assumed to be zero. The calculated value was 194 cm$^{-1}$. This value confirms the presence of a tetragonal phase, as predicted in Ref. 21, but in nanostructured FeSb$_2$ powder its nucleation was observed for pressures smaller than 38 GPa.

Figure 9 shows that, as the pressure is decreased, the tetragonal phase is observed only for pressures larger than 22.9 GPa, while the orthorhombic phase is observed for smaller pressures. It is important to note that the Raman active Sb modes are not observed after the pressure is removed, suggesting that Sb was incorporated to the amorphous, orthorhombic or tetragonal phase.

In order to analyze the Raman spectra, they were deconvoluted using Lorentzian functions, as shown in Fig. 8. For the orthorhombic and tetragonal phases, the pressure dependence of the wave number may be approximated by a standard second order polynomial, as shown in Fig. 10.

Orthorhombic phase
$$\varpi(P)A_g^2 = 159.279 + 2.082P - 0.008P^2$$
$$\varpi(P)B_{1g}^1 = 183.622 + 2.363P + 0.038P^2 \quad (6)$$
Tetragonal phase
$$\varpi(P)E_g^2 = 165.003 + 3.817P - 0.028P^2$$

The effect of high pressure on the Raman active modes can be better understood by considering the derivative of Eq. (6): $d\varpi/dP = A + 2BP$. Fig. 11 shows the derivative of analytical expressions (6) obtained from fits for $A_g^2$ and $B_{1g}^1$ modes. One can see that the derivative of Raman active $B_{1g}$ mode varies faster than the $A_g$ mode, indicating that the $B_{1g}$ mode is more sensitive to the pressure effect, as shown in Fig. 8.

The Grüneisen parameter $\gamma_0$ describes the effect of high pressure on the volume of the lattice, and, consequently, on the phonons frequencies. The zero-pressure mode Grüneisen parameters $\gamma_0$ were determined using the equation[46,47]

$$\gamma_0 = \frac{B_0}{\omega_0}\left(\frac{\partial \omega}{\partial P}\right)_{P=0} \qquad (7)$$

where $B_0$ and $\varpi_0$ are the bulk modulus in GPa and the wave number in cm$^{-1}$ at zero pressure. From the XRD measurements a value of $B_0 = 74.2 \pm 3.0$ GPa was obtained. In order to evaluate the effect of high pressure on the Raman active $A_g$ and $B_{1g}$ modes, the $\gamma_0$ value was calculated in the same pressure range (up to 12.7 GPa). Using $B_0$ and $\varpi_0$ values in Eq. (7), the $\gamma_0$ value for the $A_g$ and $B_{1g}$ modes are 0.86 and 0.94, respectively. From these values one can see that the pressure affects much more the $B_{1g}$ mode than the $A_g$ one as shown in Fig. 8. For the tetragonal phase, the $\gamma_0$ value for the Raman active $E_g^2$ mode can be estimated by considering the estimated wave number of 194 cm$^{-1}$ and $B_0 = 68$ GPa reported by Wu et al.[21] for this phase. The calculated value is 1.33.

## V. CONCLUSIONS

The structural and optical properties of nanostructured orthorhombic FeSb$_2$ phase formed by the MA technique were studied as a function of pressure, and the main conclusions are: 1) the XRD results show evidence for an orthorhombic FeSb$_2$ phase up to 28.2 GPa. 2) There is an increase in the volume fraction of an amorphous phase up to 23.3 GPa. 3) For pressures between 14.3 and 23.3 GPa, a tetragonal FeSb$_2$ phase is observed. Due to its small amount and no structural degradation of the orthorhombic phase, it is assumed that nucleation of the tetragonal phase occurred in the interfacial component; 4) The Raman active $A_g$ mode of the orthorhombic phase is observed up to the maximum pressure used and, from 21.0 to 45.2 GPa, a new Raman active $E_g^2$ mode is observed and attributed to the tetragonal phase, which was seen in the XRD measurements between 14.3 and 23.3 GPa. The wave numbers of the other Raman active modes are very close to the values of the orthorhombic phase, making difficult their assignment.

## ACKNOWLEDGMENTS

This study was supported by the Brazilian-French CAPES/COFECUB Program (Project No. 559/7), which has supported one of the authors (S.M.S). However, these data were not included in his Ph.D thesis. Thus, the analysis of the data was carried out by one of the authors (C.M.P) and the results will be part of his Ph.D thesis. We thank to the ELETTRA synchrotron (Italy) for the XRD measurements as a function of pressure. We are indebted to the high pressure group from IMPMC, in particular to Pascal Munch and Gilles Le Marchand for technical support. We are indebted to Drs. Altair Sória Pereira and Ronaldo Sérgio de Biasi for discussions and contributions.

*Present adress: Departamento de Física, Universidade Federal do Amazonas, 3000 Japiim, 69077-000 Manaus, Amazonas, Brazil.

FIGURES

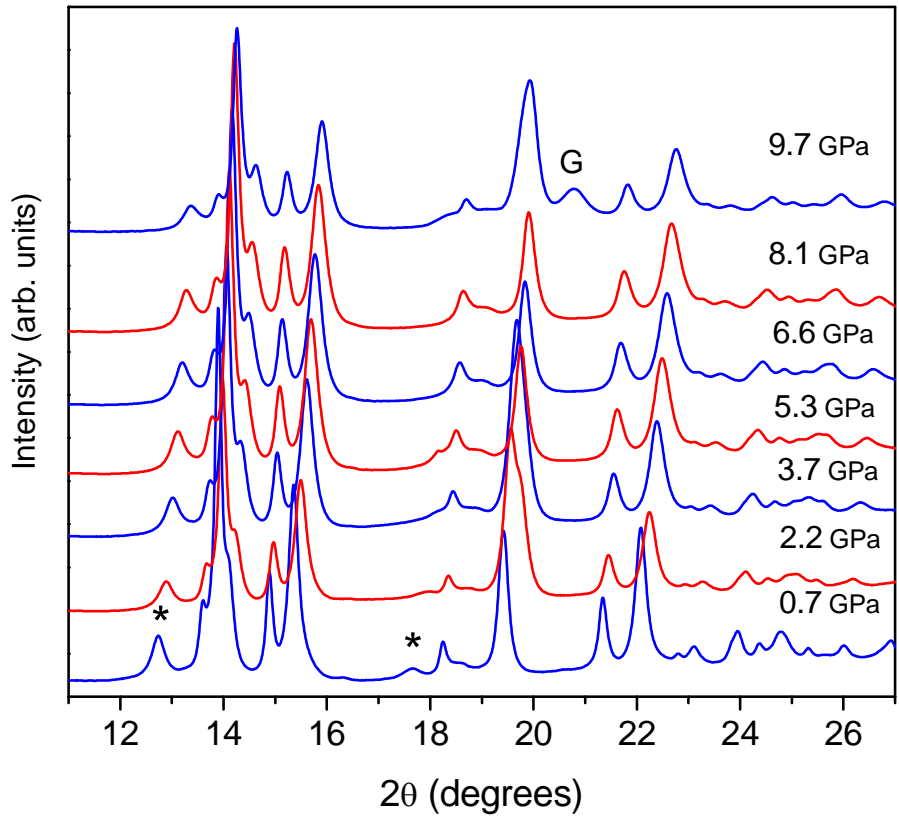

Figure 1 (color online): XRD patterns measured with increasing pressure up to 9.7 GPa for nanostructured orthorhombic $FeSb_2$ powder. The diffraction peaks of elemental Sb are marked with the asterisk (*) symbol, while those from the gasket are identified by the G letter.

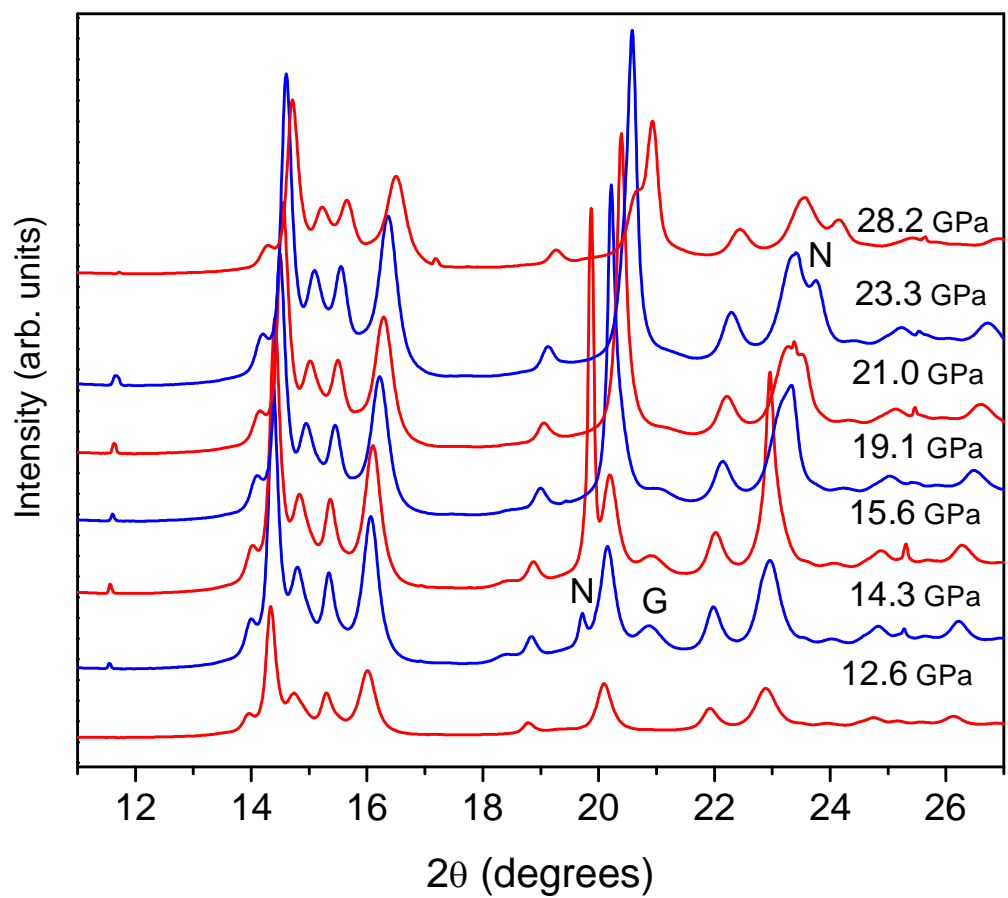

Figure 2 (color online): XRD patterns measured with increasing pressure from 12.6 up to 28.2 GPa for the nanostructured orthorhombic $FeSb_2$ powder. The diffraction peaks from the gasket and from neon are identified by the G and N letters.

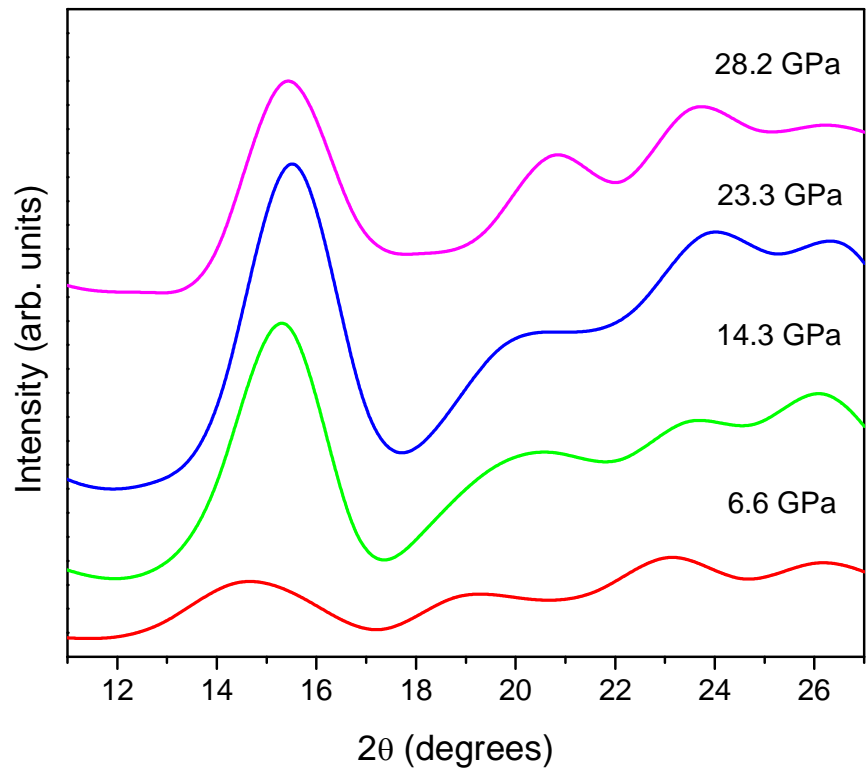

Figure 3 (color online): Estimated XRD patterns for the amorphous phase for several pressures.

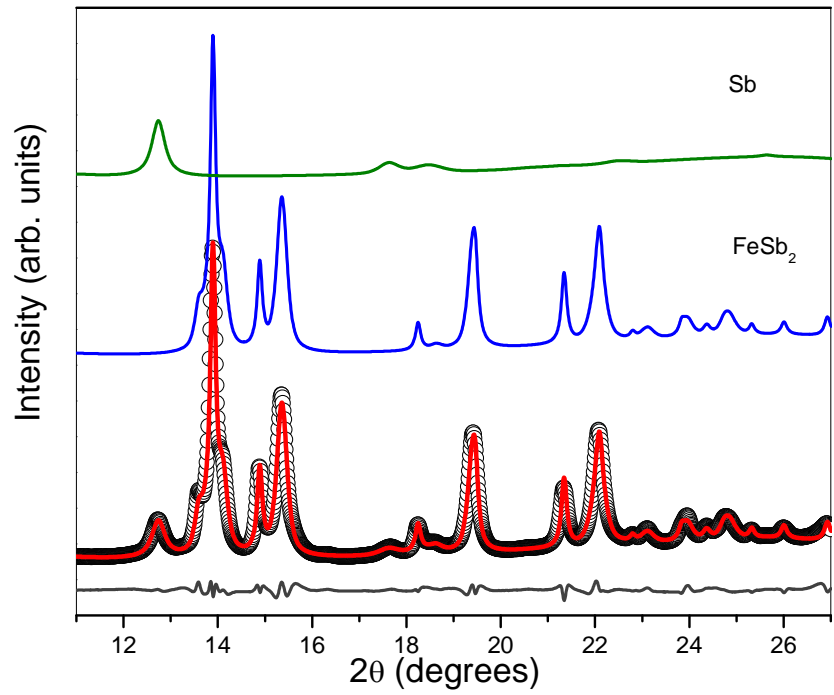

Figure 4 (color online): XRD pattern of the orthorhombic $FeSb_2$ phase at 0.7 GPa (solid black line). Other colored solid lines represent the Rietveld simulations. The bottom line is the residual intensity.

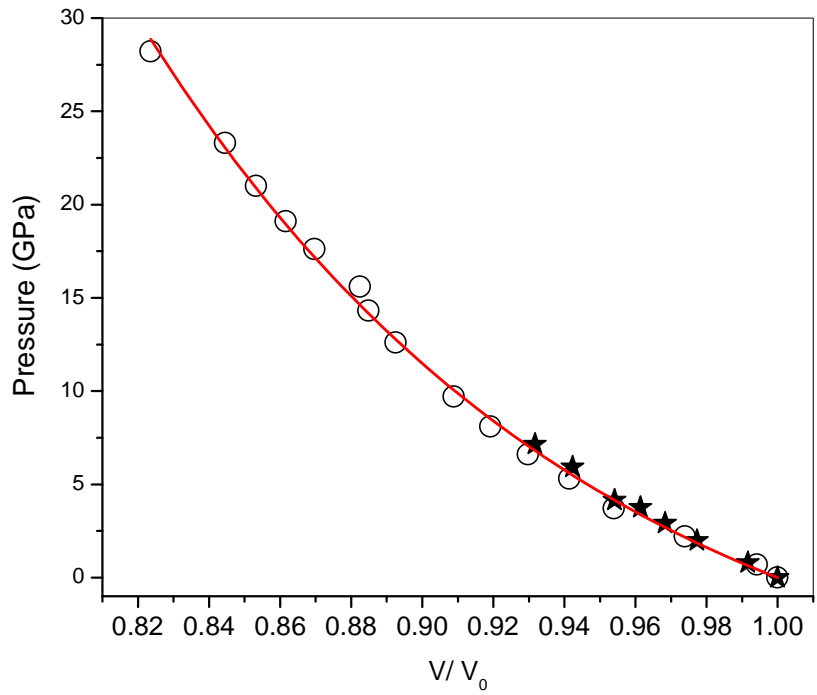

Figure 5 (color online): Pressure dependence of the volume of the nanostructured orthorhombic $FeSb_2$ phase deduced from Rietveld refinements. Present study (full circles) and results from Ref. 19 (open stars). The solid line is the fit to a Birch-Murnaghan equation of state.

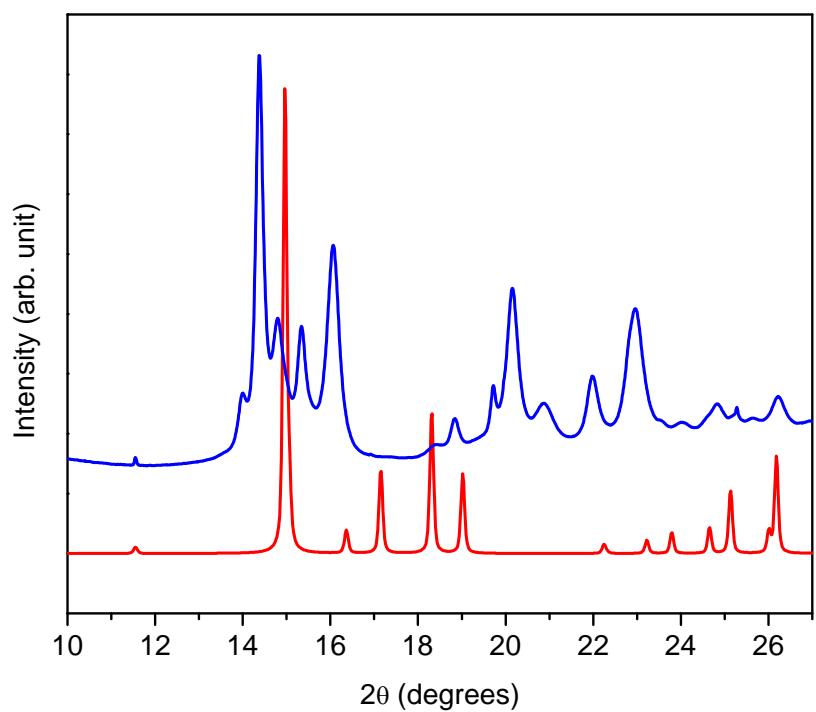

Figure 6 (color online): XRD patterns for the nanostructured FeSb$_2$ powder measured at 14.3 GPa (top curve) and simulated for the tetragonal VSb$_2$ phase using the calculated lattice parameters given in the text (bottom curve).

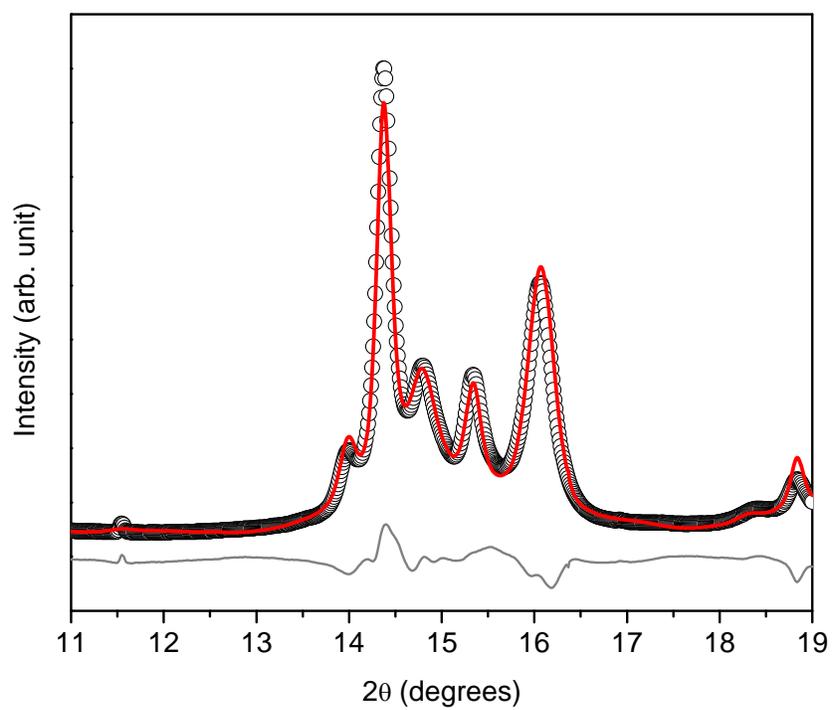

Figure 7 (color online): XRD patterns of nanostructured $FeSb_2$ powder measured at 14.3 GPa (open circles) and simulated (red solid line). The bottom line is the residual intensity.

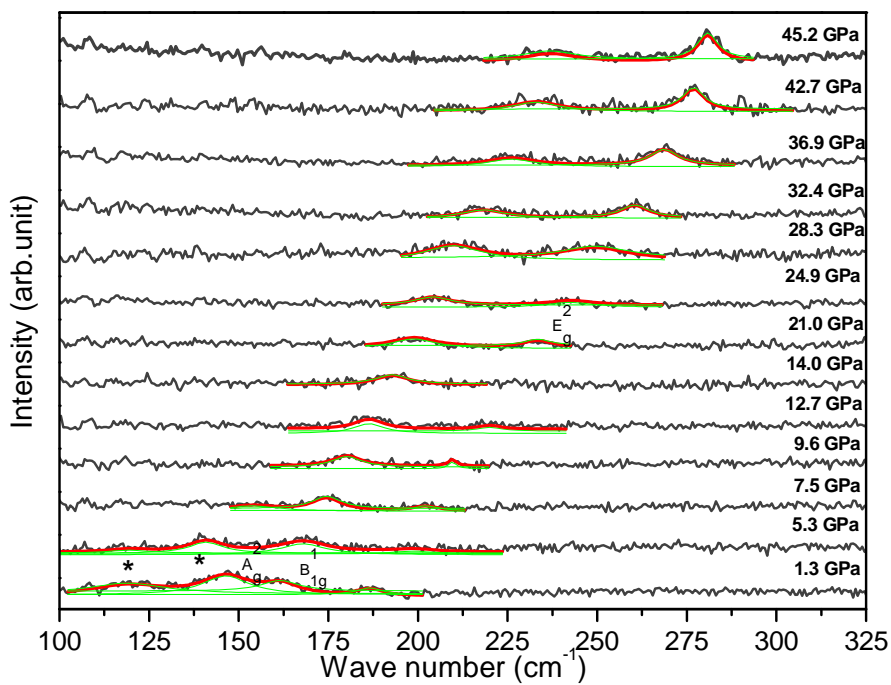

Figure 8 (color online): Raman spectra measured with increasing pressure up to 45.2 GPa for the nanostructured orthorhombic FeSb$_2$ powder. The excitation wavelength was λ = 514.5 nm. Raman active modes from unreacted antimony are marked with the asterisk (*) symbol. The red and green full lines represent the deconvolution process (see text).

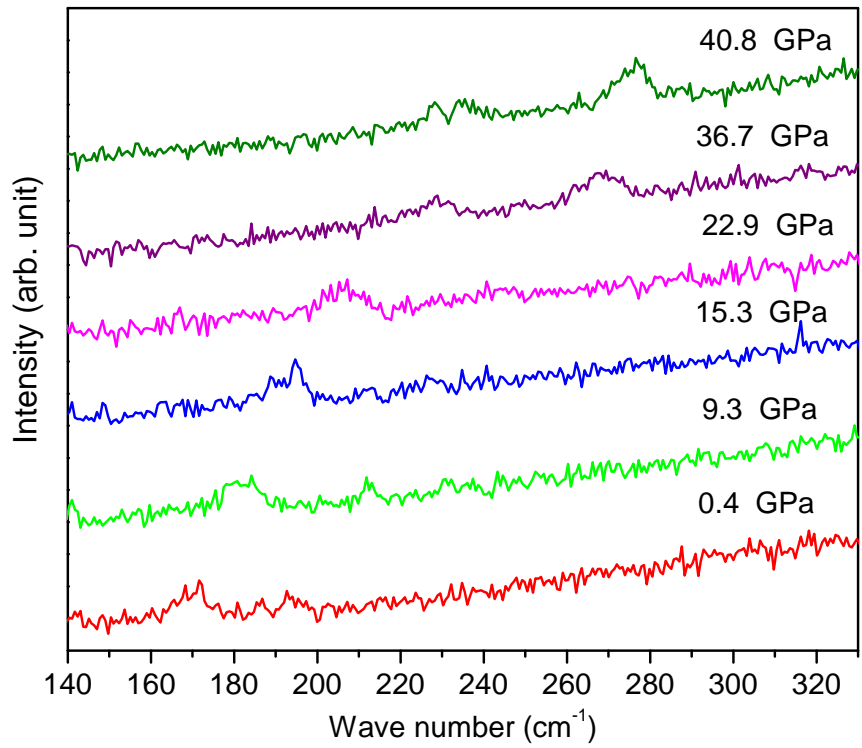

Figure 9 (color online): Raman spectra measured with decreasing pressure for the nanostructured orthorhombic $FeSb_2$ powder. The excitation wavelength was $\lambda = 514.5$ nm.

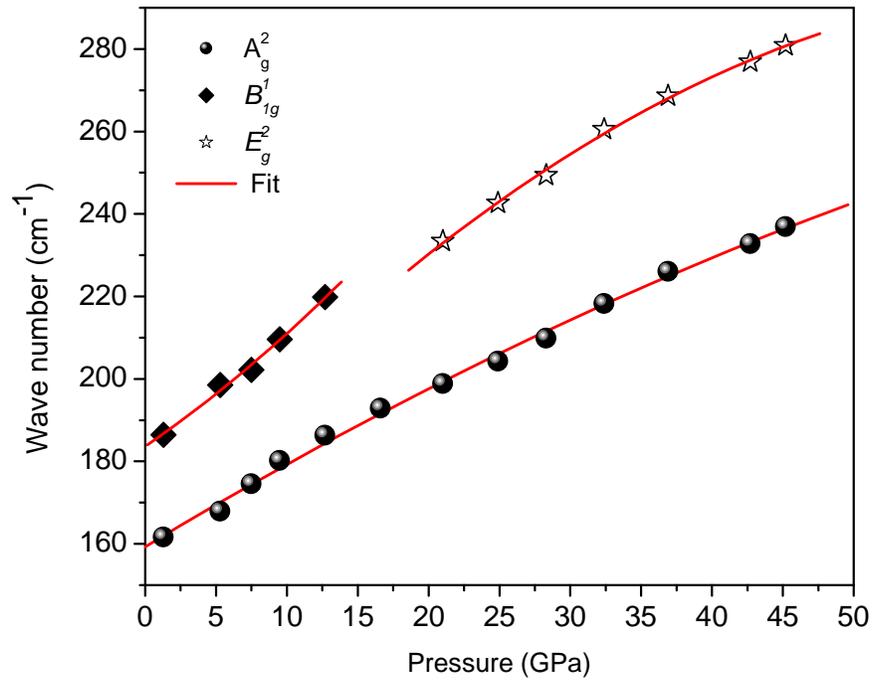

Figure 10 (color online): Pressure dependence of the Raman active modes of nanostructured orthorhombic and tetragonal FeSb$_2$ phases up to 45.2 GPa. The symbols represent the experimental data and the lines are polynomial fits (see Eq (6) in the text).

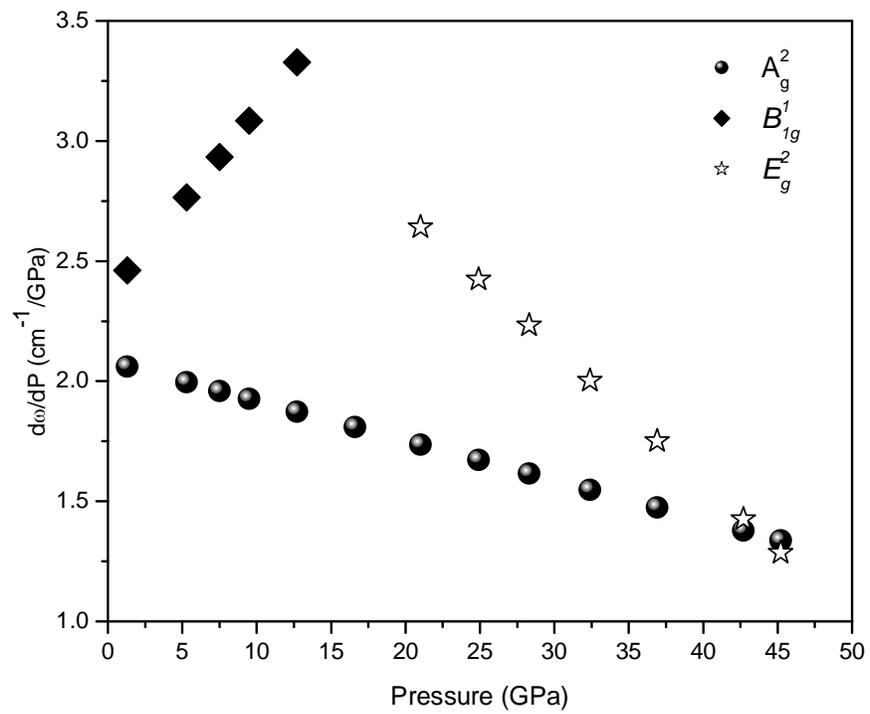

Figure 11: Derivatives of analytical expressions (6) representing the wave number values measured at several pressures.

TABLES

| P (GPa) | 0.7 | 2.2 | 3.7 | 5.3 | 6.6 | 8.1 | 9.7 | 12.6 |
|---|---|---|---|---|---|---|---|---|
| a (Å) | 5.8193 | 5.7898 | 5.7610 | 5.7440 | 5.7213 | 5.7100 | 5.6829 | 5.6698 |
| b (Å) | 6.5323 | 6.4878 | 6.4498 | 6.4233 | 6.3938 | 6.3722 | 6.3593 | 6.3175 |
| c (Å) | 3.2008 | 3.1698 | 3.1418 | 3.1237 | 3.0996 | 3.0875 | 3.0747 | 3.0553 |
| V (Å$^3$) | 121.67(3) | 119.06(7) | 116.74(0) | 115.25(0) | 113.38(6) | 112.33(9) | 111.11(7) | 109.43(7) |
| Sb (x) | 0.1878(3) | 0.1909(7) | 0.1894(8) | 0.1865(9) | 0.1897(3) | 0.1909(1) | 0.1906(1) | 0.1915(9) |
| Sb (y) | 0.3565(8) | 0.3591(7) | 0.3564(2) | 0.3548(3) | 0.3524(3) | 0.3531(7) | 0.3533(5) | 0.3447(8) |
| Fe-Sb x 2 (Å) | 2.5729(2) | 2.5582(6) | 2.5447(8) | 2.5184(5) | 2.5009(5) | 2.5007(3) | 2.4947(8) | 2.4341(3) |
| Fe-Sb x 4 (Å) | 2.5957(2) | 2.5791(6) | 2.5544(7) | 2.5592(7) | 2.5385(5) | 2.5245(4) | 2.5147(9) | 2.5204(2) |
| Fe-Sb-Fe (°) | 128.60(2) | 128.48(3) | 128.89(7) | 128.98(6) | 129.67(3) | 129.65(9) | 129.60(6) | 131.12(7) |

| P (GPa) | 14.3 | 15.6 | 17.6 | 19.1 | 21 | 23.3 | 28.2 |
|---|---|---|---|---|---|---|---|
| a (Å) | 5.6564 | 5.6501 | 5.6234 | 5.6079 | 5.5931 | 5.5776 | 5.5325 |
| b (Å) | 6.2990 | 6.2841 | 6.2708 | 6.2479 | 6.2248 | 6.2027 | 6.1577 |
| c (Å) | 3.0392 | 3.0334 | 3.0191 | 3.0100 | 2.9982 | 2.9810 | 2.9548 |
| V (Å$^3$) | 108.28(5) | 107.70(3) | 106.46(3) | 105.46(3) | 104.38(5) | 103.13(1) | 100.66(2) |
| Sb (x) | 0.1946(1) | 0.1947(1) | 0.1827(2) | 0.1952(9) | 0.1990(3) | 0.1937(8) | 0.1899(8) |
| Sb (y) | 0.3504(3) | 0.3509(5) | 0.3548(5) | 0.3497(5) | 0.3525(8) | 0.3517(7) | 0.3504(1) |
| Fe-Sb x 2 (Å) | 2.4664(2) | 2.4648(2) | 2.4506(4) | 2.4445(8) | 2.4338(1) | 2.4351(7) | 2.4001(0) |
| Fe-Sb x 4 (Å) | 2.4862(2) | 2.4804(5) | 2.5082(9) | 2.4628(1) | 2.4609(5) | 2.4461(0) | 2.4439(3) |
| Fe-Sb-Fe (°) | 130.48(6) | 130.40(4) | 128.88(7) | 130.65(4) | 130.47(2) | 130.29(0) | 130.25(6) |

Table I: Structural data for the nanostructured orthorhombic FeSb$_2$ phase.

| P (GPa) | 14.3 | 15.6 | 17.6 | 19.1 | 21 | 23.3 |
|---|---|---|---|---|---|---|
| a (Å) | 6.8441 | 6.8315 | 6.8227 | 6.8032 | 6.7985 | 6.7761 |
| c (Å) | 5.2563 | 5.2328 | 5.1902 | 5.1569 | 5.1310 | 5.1120 |
| V (Å$^3$) | 246.21(4) | 244.21(1) | 241.59(9) | 238.67(9) | 237.15(2) | 234.72(0) |
| Sb (x) | 0.1233(1) | 0.1350(2) | 0.1350(1) | 0.1349(1) | 0.1558(8) | 0.1677(2) |
| Sb (y) | 0.6233(1) | 0.6350(2) | 0.6350(1) | 0.6349(1) | 0.6558(8) | 0.6677(2) |
| Fe-Fe x 2 (Å) | 2.6281(5) | 2.6164(0) | 2.5951(0) | 2.5784(5) | 2.5655(0) | 2.5560(0) |
| Fe-Sb x 8 (Å) | 3.0142(8) | 2.9630(2) | 2.9552(5) | 2.9451(4) | 2.8707(9) | 2.8274(9) |
| Fe-Sb x 8 (Å) | 4.5428(0) | 4.6238(7) | 4.6153(9) | 4.6001(3) | 4.6265(5) | 4.5892(3) |

Table II: Structural data for the tetragonal FeSb$_2$ phase.